\documentclass[aps,prd,nofootinbib,twocolumn,amsmath,amssymb,longbibliography]{revtex4-1}

\usepackage{amsmath,amssymb,amsfonts,latexsym,cancel}

\usepackage{mathtools}
\usepackage{graphicx}
\usepackage{color}
\usepackage{amsbsy}
\usepackage{hyperref}
\usepackage{tikz}

\usepackage{seqsplit}

\usepackage[caption=false]{subfig}

\usepackage{comment}

\newcommand{\be}{\begin{equation}}
\newcommand{\beq}{\begin{equation}}
\newcommand{\ee}{\end{equation}}

\def\bea {\begin{eqnarray}}
\def\eea {\end{eqnarray}}

\def\f{\frac}

\def\nn{\nonumber}
\def\lp{\ell_{\rm Pl}}

\def\bra{\langle}
\def\ket{\rangle}

\def\mP{\mathcal{P}}

\begin{document}

\title{Tunnelling across a trapped region and out of a black hole}

\author{Edward Wilson-Ewing} \email{edward.wilson-ewing@unb.ca}
\affiliation{Department of Mathematics and Statistics, University of New Brunswick, \\
Fredericton, NB, Canada E3B 5A3 \looseness=-1}

\begin{abstract}

The quantum field theory for a massless scalar field on a two-dimensional non-singular black hole spacetime with an outer and an inner horizon predicts that there is a non-vanishing probability for a particle localized inside the inner horizon to tunnel out of the black hole. The tunnelling probability monotonically asymptotes to a maximal value that is inversely proportional to the sum of the inverse surface gravities of the inner and outer horizons of the black hole.

\end{abstract}

\maketitle

\section{Introduction}

In quantum mechanics, it is possible for a particle to tunnel through a potential barrier even if the particle has less energy than the height of the potential walls. This is a quantum phenomenon, with no classical equivalent.

Tunnelling can be understood more generally as any process that is classically impossible, but that has a non-zero probability of occurring in the underlying quantum theory. Another example arises in quantum field theory, where particles can tunnel outside their lightcone. One measure of this is given by the propagator between two spacelike-separated points; for a massless particle, the propagator scales as $1/d^2$ in 3+1 dimensions and as $\ln(1/d)$ in 1+1 dimensions, where $d$ is the (spacelike) distance between the two points.

This can be made more precise by calculating transition amplitudes between single-particle states that are localized in regions that are causally disconnected: it is possible for a particle to tunnel between the causally-disconnected regions \cite{Hegerfeldt:1974qu} (this does not imply any violations of causality \cite{Eberhard:1988yj}), and the tunnelling probability decays polynomially (not exponentially) with $d$.

Further, Hawking radiation \cite{Hawking:1975vcx}, the quintessential effect of quantum field theory on a curved but fixed background spacetime, can also be described as a tunnelling process \cite{Srinivasan:1998ty, Volovik:1999fc, Parikh:1999mf}, where the Hawking quanta tunnel a short distance across the horizon; for a broad introduction to quantum field theory in curved spacetime see, e.g., \cite{Birrell:1982ix}.

These results raise an obvious question: is a different type of tunnelling process possible, where a particle would tunnel out of a black hole from its centre?

For most black hole solutions in general relativity (including the astrophysically relevant rotating black holes), the trapped region is bounded on the exterior by an outer horizon, and also on the interior by an inner horizon: there exists an inner region lying inside the inner horizon where particles can move both inwards and outwards. Further, inner horizons may be present also in (uncharged) non-rotating black holes as well. Although the classical Schwarzschild solution does not contain an inner horizon, it does contain a curvature singularity at its centre, and it is generally expected that quantum gravity effects will resolve the singularity. If the quantum-corrected spacetime is indeed singularity-free, and if in addition quantum fluctuations of the spacetime geometry are sufficiently small so there exists a metric that captures to a reasonable degree of accuracy the spacetime geometry everywhere, then this metric will generically contain an inner horizon in addition to the outer horizon \cite{Hayward:2005gi}; for recent reviews of non-singular black hole models see, e.g., \cite{Malafarina:2017csn, Maeda:2021jdc}. The presence of an inner horizon even in spherical symmetry is not surprising: in collapse models (for example~\cite{Oppenheimer:1939ue}), inner and outer horizons form in pairs, and the only reason there is no inner horizon at the end of a classical collapse is that the inner horizon has been swallowed by the singularity at the centre of the black hole---if the singularity is resolved by quantum gravity effects, the inner horizon will survive. (A possible exception is if the black hole has a region with large quantum fluctuations and no effective metric, then this quantum region could swallow the inner horizon, like the singularity does in the classical theory.)

Inner horizons are typically subject to mass inflation, an instability due to gravitational blueshift at the inner horizon \cite{Poisson:1990eh, Ori:1991zz}. (Contrary to eternal black holes, there is no Cauchy horizon at the inner horizon in a black hole that forms from gravitational collapse, but mass inflation occurs nonetheless \cite{Carballo-Rubio:2024dca}.) In some cases, mass inflation can be avoided, for example if the inner horizon is extremal \cite{Carballo-Rubio:2022kad}, in Eddington-inspired Born-Infeld gravity \cite{Avelino:2015fve}, and in some loop quantum gravity effective models \cite{Husain:2021ojz}, but otherwise its presence is quite robust. Despite this, the fact binary black holes survive long enough to slowly spiral inwards and merge \cite{LIGOScientific:2016aoc} shows that mass inflation cannot drastically shorten the lifetime of black holes and, assuming black holes remain non-singular, the inner horizon will survive if the core has a regular geometry and the null expansion is everywhere continuous \cite{Hayward:2005gi}. Therefore, in the following I assume that while non-linearities may be large (and the exact physics, perhaps due to quantum gravity, may remain unclear), the inner horizon survives mass inflation.

So, for a non-singular black hole with both an outer and an inner horizon, is it possible for a particle located inside the inner horizon to tunnel across the trapped region and out of the black hole?  Classically such a process is forbidden but, as shall be shown here, quantum field theory makes this possible.

\section{Background Geometry}

For the sake of simplicity, consider a two-dimensional spacetime with a trapped region bounded by an outer horizon and an inner horizon with the metric, in Painlev\'e-Gullstrand coordinates,
\beq \label{metric-pg}
ds^2 = -dt^2 + (dx - v \, dt)^2,
\ee
where $v(x)$ depends on $x$ but, again for simplicity, is independent of $t$. Following the usual convention, I take $v \le 0$, and for the spacetime to contain a single trapped region, I require that $v=-1$ at exactly two values of $x$, the inner and outer horizons located at $x_i$ and $x_o$ respectively, with $x_i < x_o$. I also assume that $v(x)$ is everywhere a smooth function so there are no curvature singularities, and that $dv/dx$ is non-zero at the horizons so the horizons are not extremal.

Since this spacetime is non-singular and contains a trapped region, it provides a simple two-dimensional model for a non-singular black hole. Two-dimensional modes like this are very useful because, on the one hand, they can capture much of the relevant physics, and on the other hand, calculations are often considerably simpler, especially for conformally invariant matter since two-dimensional spacetimes are conformally flat. Indeed, Hawking radiation and Unruh radiation can both be studied to great effect in two dimensions \cite{Unruh:1976db} (although one difference is the lack of greybody factors for Hawking radiation in two dimensions), and backreaction calculations are also simpler in two dimensions \cite{Callan:1992rs, Fabbri:2005mw, Ashtekar:2010hx, Barenboim:2024dko}.

The spacetime with the metric \eqref{metric-pg} contains three regions: the outer region $x > x_o$, the trapped region $x_i < x < x_o$, and the inner region $x < x_i$; the Painlev\'e-Gullstrand coordinates are convenient in that they cover all three regions without any coordinate singularities. Classically, causal test particles can move inwards (towards smaller $x$) and outwards in the outer and inner regions, but they must move inwards in the trapped region. (While classical particles in the inner region can move outwards, they can only asymptotically approach $x_i$ and can never enter the trapped region, let alone reach the outer region.) The black hole region is the set of all points that cannot escape to positive infinity; this is the union of the trapped and inner regions.

Although this metric is static with the time-like Killing vector $\xi^a = (\partial / \partial t)^a$, this is an approximation. Rather, the trapped region should be understood to have formed dynamically in the past from gravitational collapse, and will eventually evaporate in the future. The assumption here is the black hole is approximately static for a long time between its initial formation and eventual evaporation, during which time the metric \eqref{metric-pg} is a good approximation. Since the black hole is not eternal, there is no maximal extension to the spacetime.

It is helpful to briefly review the geodesics of this spacetime. Time-like geodesics $u^a = (dt/d\tau, dx/d\tau)$, where $\tau$ is the proper time along the geodesic,  satisfy $(dx/d\tau)^2 = E^2-1-v^2$, where $E = -\xi_a u^a$ is the energy per unit rest mass of a particle following the geodesic as measured by a static observer at infinity. The case $E=1$ corresponds to a particle at rest at infinity, and then $dx/d\tau = v$, and also $\tau=t$; since $v \le 0$ the geodesic moves from right to left.

For null curves, $ds^2=0$ and then $dx/dt = v \pm 1$.  If $v < -1$, then even outgoing null rays will move inwards so regions where $v < -1$ are trapped, with Killing horizons located where $v=-1$.

It is convenient to introduce null coordinates to describe this spacetime, as a first step define
\beq
x_\pm = \int \f{dx}{1 \pm v}.
\ee
While the coordinate $x_-$ covers all three (inner, trapped, outer) regions, the coordinate $x_+$ diverges at both $x_i$ and $x_o$ where $v=-1$, and because of this $x_+$ can cover any one of the regions (but not all three at once); this coordinate in each of the regions is denoted by $x_{(i)+}, x_{(t)+}, x_{(o)+}$ with the additional subscripts corresponding to the inner, trapped, and outer regions respectively.

Then, in terms of the null coordinates
\beq
u = t-x_+, \qquad V = t + x_-,
\ee
the metric is simply $ds^2 = -(1-v^2) \, du \, dV$. Like the $x_\pm$ coordinates, while the $V$ coordinate covers all three regions, the $u$ coordinate only covers a single region at a time, diverging at both the inner and outer horizons. Because of this, there are three null coordinates $u$, namely $u_{(i)}, u_{(t)}, u_{(o)}$, that each cover one of the three regions (note that $u_{(o)}, u_{(t)} \to +\infty$ at the outer horizon, and that $u_{(i)}, u_{(t)} \to -\infty$ at the inner horizon).

To obtain a null coordinate $U$ that (like $V$) covers the three regions simultaneously, it is necessary to find a coordinate transformation $U = U(u)$ such that all three of $u_{(i)}, u_{(t)}, u_{(o)}$ are covered by $U$. Further, to avoid coordinate singularities, the relation $dU = f(U) du$ must be such that the conformal factor in the resulting metric $ds^2 = -[(1-v^2)/f(U)] \, dU \, dV$ never vanish or diverge: therefore, $f(U)$ must have the same zeros as $(1+v)$ and these two quantities must also go to zero at the same rate at the horizons so $(1+v)/f(U)$ remains finite. To impose this last condition, note the Taylor expansion for $v$ is
\beq \label{taylor-outer}
v = -1 + \alpha_o (x-x_o)
\ee
near the outer horizon (with $\alpha_o > 0$), and
\beq
v = -1 - \alpha_i (x-x_i)
\ee
near the inner horizon (with $\alpha_i > 0$); $\alpha_o$ and $\alpha_i$ are the surface gravities of the horizons. Then,
\beq \label{def-U}
\!\!\! U \!=\! \begin{cases}
\!-\alpha_o^{-1} (e^{-\alpha_o u_{(o)}} +1), \!
&{\rm outer~region}, \\
\alpha_o^{-1} (e^{-\alpha_o u_{(t)}} - 1),
&{\rm trapped~region}, u_{(t)} > 0, \\
\!-\alpha_i^{-1} (e^{\alpha_i u_{(t)}} -1),
&{\rm trapped~region}, u_{(t)} < 0, \\
\alpha_i^{-1} (e^{\alpha_i u_{(i)}} + 1),
&{\rm inner~region},
\end{cases}
\ee
satisfies the required conditions, and $dU = f(U) du$ with
\beq
f(U) = \begin{cases}
-1 -\alpha_o U, \: & U < 0, \\
-1 + \alpha_i U, & U > 0.
\end{cases}
\ee
Note that $U < -1/\alpha_o$ in the outer region, $-1/\alpha_o < U < 1/\alpha_i$ in the trapped region, and $U > 1/\alpha_i$ in the inner region; $U=-1/\alpha_o$ at the outer horizon and $U=1/\alpha_i$ at the inner horizon. Importantly, $f(U)$ is negative in the trapped region, and further $(1+v)/f(U)$ remains finite at the horizons, for example near the outer horizon where the Taylor expansion \eqref{taylor-outer} holds, $(1+v)/f(U) = e^{\alpha_o t}$. The one drawback for this choice of $U$ is that at $U=0$ it is a once-differentiable function of $u$, since $f(U)$ is continuous but not differentiable at $u_{(t)} = 0$. By changing the definition for $U$ in a small neighbourhood of $U=0$, $U(u)$ can be made twice-differentiable (or even smooth) everywhere, but this will not affect the resulting physics; for simplicity I will work with \eqref{def-U} in the following.

The metric in terms of the null coordinates $U,V$ is
\beq \label{metric-UV}
ds^2 = - \, \f{(1-v^2)}{f(U)} dU \, dV.
\ee
This form of the metric is explicitly conformally flat, with a conformal factor that is everywhere strictly positive, and it covers the outer, trapped and inner regions.

It is also convenient to introduce
\beq
T = \f{U+V}{2}, \quad X = \f{V-U}{2},
\ee
$T$ is everywhere timelike, and $X$ is everywhere spacelike.

\section{Quantum Field}

Now, consider a test massless scalar field $\phi$ living on this spacetime with action
\beq
S = -\f{1}{2} \int d^2x \sqrt{-g} \, g^{ab} (\partial_a \phi) (\partial_b \phi).
\ee
This scalar field has the advantage of being conformally invariant, so the Klein-Gordon equation for $\phi$ on the space-time with the metric \eqref{metric-UV} is the same as on Minkowski space, namely $\partial_U \partial_V \phi = 0$, or equivalently
\beq
- \partial_T^2 \phi + \partial_X^2 \phi = 0,
\ee
with the solutions being the usual linear combinations of the plane waves $e^{-ik(T \pm X)}$.

The field operator $\hat\phi(X,T)$ in the corresponding quantum field theory, expressed in terms of the plane-wave basis of the classical solutions in $X,T$ coordinates, is
\beq
\hat \phi = \f{1}{2\pi} \! \int_{-\infty}^\infty \! \f{dk}{\sqrt{2|k|}} \Big( e^{-i|k|T + ikX} a_k + e^{i|k|T - ikX} a_k^\dag \Big),
\ee
where the creation and annihilation operators satisfy the usual commutation relations $[a_k, a_p^\dag] = 2\pi \delta(k-p)$ and $[a_k, a_p] = [a_k^\dag, a_p^\dag] = 0$.

In terms of these coordinates, the momentum conjugate to the scalar field is $\pi_\phi = \partial \phi/\partial T$, so the momentum operator is $\hat \pi_\phi(X,T) = \partial_T \hat\phi(X,T)$, just as in Minkowski space. The calculation for the Hamiltonian is again the standard one, with the classical Hamiltonian being $H = \tfrac{1}{2} \int dX (\pi_\phi^2 + (\partial_X \phi)^2)$, and the Hamiltonian operator is (after normal-ordering as usual)
\beq \label{qft-ham}
\hat H = \f{1}{2\pi} \int_{-\infty}^\infty dk \, |k| a_k^\dag a_k.
\ee
Finally, the Fock vacuum $|0\ket$ is defined by requiring that $a_k |0\ket = 0$ for all annihilation operators $a_k$. As usual for quantum field theory on a curved background, this is the ground state for observers whose proper time is $T$.

Before studying tunnelling, the propagator, or two-point correlation function, for $\phi$ can give some intuition. In two dimensions, the propagator diverges in the infrared and must be regulated. Ignoring its regulator-dependent amplitude, the propagator scales as \cite{DiFrancesco:1997nk}
\beq \label{corr}
\bra 0 | \hat \phi(X_2,T_2) \hat \phi(X_1, T_1) | 0 \ket \sim \ln(1/\Delta s^2),
\ee
where $\Delta s^2 = - (T_2 - T_1)^2 + (X_2 - X_1)^2$, and this is non-vanishing even when $\Delta s^2 > 0$. This result does not violate causality: in quantum field theory, causality requires that operators evaluated at space-like separated points commute, and that condition is satisfied here. Rather, this demonstrates the fact that the vacuum state $|0\ket$ contains correlations between points that are spacelike-separated, since $|0\ket$ is defined as the state that is annihilated by the non-local operators $a_k$ (these correlations are also important for entanglement entropy calculations \cite{Ribes-Metidieri:2025nfw}). While this equation for the propagator provides some physical intuition for the tunnelling process studied below, it will not be used in the following.

The fact the propagator is non-zero even between points that are causally disconnected---including between points inside and outside the black hole---suggests it may be possible for a particle to tunnel between causally disconnected regions, including out of the black hole.

To determine whether this occurs, consider 1-particle states of the form
\beq
|\psi\ket = \f{1}{2\pi} \int_{-\infty}^\infty dk \, \tilde \psi(k) e^{i|k| T_0} a_k^\dag |0\ket.
\ee
These states correspond to single-particle states with a wavepacket $\psi(X)$ at time $T_0$, and $\tilde\psi(k)$ is the Fourier transform of $\psi(X)$. Here I will consider two simple types of localized (at position $X=X_\ell$) states: the first a delta function localization $\psi(X) = \delta(X-X_\ell)$, and the second a flat wave packet of width $W$ where $\psi(X) = 1/\sqrt{W}$ for $|X-X_\ell| < W/2$ and $\psi(X)=0$ elsewhere. These two states are, respectively,
\beq
|X_\ell, T_0\ket_\delta^{} = \f{1}{2\pi} \int_{-\infty}^\infty dk \, e^{i|k| T_0 - i k X_\ell} a_k^\dag |0\ket,
\ee
\beq
\!\!\! |X_\ell, T_0\ket_W^{} \!=\! \f{1}{\pi\sqrt{W}} \int_{-\infty}^\infty \!\!\f{dk}{k} \sin\!\f{Wk}{2} \, e^{i|k| T_0 - i k X_\ell} a_k^\dag |0\ket.
\ee
While the delta-function localized state has a simpler form, that state is not normalizable.

For the delta-function localized states,
\beq
{}_\delta^{}\bra X_2, T_2 | X_1, T_1 \ket_\delta^{} = \f{i \Delta T}{\pi \Delta s^2}.
\ee
When $\Delta s^2 > 0$, the non-vanishing overlap between an initial and a final state for a single particle with a delta-function localization at two points that are space-like separated implies a non-zero probability for a process that is classically forbidden, and can be understood as quantum tunnelling.  (In fact, for a massless field like $\phi$, any $\Delta s^2 \neq 0$ corresponds to a classically forbidden process, but the more interesting case is space-like propagation, which is impossible both for massless and massive particles.) There are two other points to note here. First, the divergence at $\Delta s = 0$ is due to working with a non-normalizable state; for a normalized state propagation on the lightcone gives a large but finite transition amplitude, with a corresponding probability less than 1. Second, while the theory is Lorentz invariant, this transition amplitude is not because the states $|X_\ell,T\ket$ localize particles at an instant of time, in a particular frame. These states are not Lorentz invariant, so the transition amplitude between the states is not Lorentz invariant either.

The states $|X, T\ket_\delta^{}$ also provide a basis for 1-particle states, and therefore for any 1-particle state $|1\ket$ the probability density that at time $T$ the particle be found at position $X$ is simply $|{}_\delta^{}\bra X, T | 1 \ket|^2$. Therefore, the probability $P_{out}$ to find a particle in the outer region (where $X \ge X_o = T + \alpha_o^{-1}$) is
\beq
P_{out} = \int_{X_o}^\infty dX \, \Big| {}_\delta^{}\bra X,T| 1\ket \Big|^2.
\ee

Now, consider the 1-particle state where the particle is localized with a flat wave packet of width $W$ just inside the inner horizon. Clearly, the probability for the particle to tunnel outside the black hole at time $T_2 \ge T_1$ is
\beq
P_{out} = \int_{X_o}^\infty dX \, \Big| {}_\delta^{}\bra X,T_2| X_\ell,T_1\ket_W^{} \Big|^2,
\ee
where for simplicity I take $X_\ell = X_i-W/2$ so the centre of the initial wave packet of width $W$ lies a distance $W/2$ inside the inner horizon.

Using the approximation $W \ll \Delta X \pm \Delta T$ (for both signs), in this limit
\beq \label{transition-Wd}
{}_\delta^{}\bra X_2, T_2 | X_1, T_1 \ket_W^{} = \f{i \sqrt{W} \Delta T}{\pi \Delta s^2},
\ee
so that
\beq
P_{out} = \int_{\Delta X_o}^\infty d(\Delta X) \, \f{W \Delta T^2}{\pi^2 (\Delta X^2 - \Delta T^2)^2},
\ee
where $\Delta X_o = \Delta T+\alpha_o^{-1}+\alpha_i^{-1}$ (neglecting the contribution of $W$ to the lower bound of the integral), with the result
\begin{align} \label{prob-out}
P_{out} =& \, \f{W}{\pi^2} \Bigg( \f{\Delta T + \alpha_i^{-1} + \alpha_o^{-1}}{4 \Delta T(\alpha_i^{-1} + \alpha_o^{-1}) + 2(\alpha_i^{-1} + \alpha_o^{-1})^2} \nn \\ &
- \f{1}{4\Delta T} \ln\left( \f{2\Delta T + \alpha_i^{-1} + \alpha_o^{-1}}{\alpha_i^{-1} + \alpha_o^{-1}} \right) \Bigg).
\end{align}
This monotonically asymptotes as $\Delta T \to \infty$ to
\beq \label{asym}
{\mP}_{out} = \f{W}{4\pi^2 (\alpha_i^{-1} + \alpha_o^{-1})}.
\ee
As stated above \eqref{transition-Wd}, this calculation assumes $W \ll \Delta X \pm \Delta T$. Since the lower bound for $\Delta X$ is $\Delta X_o = \Delta T+\alpha_o^{-1}+\alpha_i^{-1}$, this result holds only when $W \ll \alpha_o^{-1}+\alpha_i^{-1}$; for larger $W$ the calculation has the same steps but the inner product \eqref{transition-Wd} must be computed exactly. As expected, even at its maximal asymptotic value ${\mP}_{out} \ll 1$.

For finite $\Delta T$, departures from the asymptotic value of $\mP_{out}$ are to leading order $\sim W \Delta T^{-1} \ln\big((\alpha_i^{-1} + \alpha_o^{-1})/\Delta T\big)$.
The tunnelling probability increases with $W$, and can also be amplified by increasing the smaller of the two surface gravities. Interestingly, the asymptotic value of $\mP_{out}$ depends on the background spacetime geometry only through the surface gravities $\alpha_i$ and $\alpha_o$.

Finally, another way to amplify this effect is to create $N$ excitations localized at $X_1, T_1$,
\beq
\!\! |X_1^N, T_1^N\ket_\delta^{} \!=\! \f{1}{\sqrt{N!}} \! \left( \f{1}{2\pi} \! \int_{-\infty}^\infty dk \, e^{i|k|T_1 - i k X_1} a_k^\dag \right)^{\!\!N}  \!\! |0\ket,
\ee
and then the transition amplitude for one of the excitations to tunnel is amplified by a factor of $\sqrt N$:
\beq
\!\!\!\!\! {}_{\delta\!}^{} \bra X_1^{N\!-\!1}\!\!, T_1^{N\!-\!1}; X_2,\! T_2| X_1^N \!\!, T_1^N\ket_{\!\delta}^{} \! = \! \sqrt{\! N} \, {}_{\delta\!}^{}\bra X_2,\! T_2 | X_1,\! T_1 \ket_{\!\delta}^{}, \!\!\!\!
\ee
with a similar definition and result for $|X_1^N, T_1^N\ket_W$.
Therefore, the probability for any one of the (indistinguishable) particles to tunnel out, from the inner to the outer region, is increased by a factor of $N$. (For $N>1$, there is also the possibility that two or more particles tunnel out but, unless $N$ is sufficiently large, the probability of such a process is subleading compared to the 1-particle tunnelling probability.)

\section{Discussion}

The probability to tunnel across a trapped region and exit the black hole is small but non-vanishing, and is only polynomially suppressed by $\alpha_i^{-1} + \alpha_o^{-1}$. While the calculations here concern a massless scalar field in a two-dimensional spacetime, the same effect arises more generally including in four-dimensional non-singular black hole spacetimes.

While this effect also occurs in four dimensions, there are some differences since then a massless scalar field is not conformally invariant and there are greybody factors. Further, the tunnelling probability scales differently in four dimensions. In Minkowski space an analogous calculation is to compute the probability for a particle localized up to some width $W$ to tunnel outside an outgoing null surface that lies a distance $L$ away at the initial time. The tunnelling probability for this process asymptotes to a maximum that scales as $\sim W/L$ in 1+1 dimensions (matching $\mP_{out}$ in \eqref{asym} with $L$ replaced by $\alpha_i^{-1}+\alpha_o^{-1}$), while the scaling is $\sim (W/L)^3$ in 3+1 dimensions.  Given the close correspondence between black holes and null surfaces in Minkowski space \cite{DeLorenzo:2017tgx}, it seems likely that the tunnelling probability for a four-dimensional black hole also asymptotes to $\sim (W/(\alpha_i^{-1}+\alpha_o^{-1}))^3$, although this remains to be verified.

The tunnelling effect studied here (which can only occur for excited states) is distinct from Hawking radiation where the vacuum state $|0\ket$ defined below \eqref{qft-ham} appears thermal to an observer who only has access to the exterior region. What would such an external observer see if, instead of the vacuum state, there is a 1-particle state localized inside the inner horizon? The usual Bogolyubov transformation applied to (the expectation value of) the number operator acting on such a state simply gives the usual Hawking thermal bath, plus an additional contribution due to the particle localized inside the inner horizon.

An important question is whether this tunnelling process, that allows particles inside the black hole to escape while respecting unitarity, may resolve the information loss problem. To some extent, a non-singular black hole spacetime (with a single asymptotic region, to avoid information escaping to a different asymptotic region) avoids the information loss problem since there is no singularity to swallow the information \cite{Hayward:2005gi}. Still, the question remains of how to purify the Hawking radiation.

Whether tunnelling can purify Hawking radiation depends on the rate at which the tunnelling occurs. To get an order-of-magnitude estimate of the rate of tunnelling for a spherically symmetric vacuum black hole, assume that: (i) there is backreaction where the black hole loses mass for each portion of each particle's wave packet that lies beyond the outer horizon; (ii) within a time $T_{asym} \sim \alpha_i^{-1}+\alpha_o^{-1}$, the tunnelling probability nears its asymptotic maximum for a four-dimensional black hole of $\mP_{out} \sim (W/(\alpha_i^{-1}+\alpha_o^{-1}))^3$, and the classical value $\alpha_o^{-1} = 4M$ holds for the outer horizon; (iii) there is an inner horizon due to quantum gravity effects, so $\alpha_i^{-1} \sim \lp \ll \alpha_o^{-1}$, and the inner region extends from the origin to the region where, classically, the Kretschmann scalar $K = R_{abcd} R^{abcd}$ reaches the Planck scale, so $K \sim M^2/r^6 \sim \lp^{-4} \Rightarrow r_i \sim (M\lp^2)^{1/3}$, as occurs for some black hole models in loop quantum gravity \cite{Gambini:2020nsf, Kelly:2020uwj, Lewandowski:2022zce, Han:2023wxg}; (iv) the width for the wave packet of the particles is of the order of the radius of the inner region, $W \sim (M \lp^2)^{1/3}$; and (v) the black hole contains $N \sim M$ particles that are the source of the gravitational field, so there are $\sim N$ particles that can tunnel out. Then,
\beq
\f{dM}{dT} \sim - \f{N \mP_{out}}{T_{asym}} \sim - M^{-2}.
\ee
Further, close to the horizon, where most tunnelling occurs, $T\sim t$ up to prefactors of order 1, so $dM/dt \sim -M^{-2}$. (There is an ambiguity in the choice of $T$, since Lorentz transformations in $X,T$ leave the metric conformally flat; here I assume a time dilation factor of order 1.) Since $dM/dt \sim -M^{-2}$ is the same rate as Hawking evaporation, tunnelling may be a comparably important effect and perhaps could help purify Hawking radiation, for example by allowing the negative-energy Hawking pair that fell into the black hole to tunnel out and purify the emitted Hawking quantum. A more detailed exploration of the impact of this tunnelling effect on Hawking radiation and the information loss problem is left for future work.

Finally, since the prospect of experimental tests of quantum effects in black holes remain remote, it is interesting to consider analogue models that capture the same types of effects and are more easily accessible experimentally \cite{Unruh:1980cg, Barcelo:2005fc}. In particular, tunnelling across a trapped region can also occur in analogue systems.

\medskip

\noindent
{\it Acknowledgments:}
I thank Keshav Dasgupta and Viqar Husain for helpful discussions, 
and the McGill University Physics Department for their hospitality during the final stages of this work.
This work was supported in part by the Natural Sciences and Engineering Research Council of Canada.

\raggedright

\end{document}